# Total Mass of Ordinary Chondrite Matter Originally Present in the Solar System

J. Marvin Herndon

Transdyne Corporation
San Diego, California 92131 USA

October 8, 2004

## Abstract

Recently, I reported the discovery of a new fundamental relationship of the major elements (Fe, Mg, Si) of chondrites that admits the possibility that ordinary chondrite meteorites are derived from two components, a relatively oxidized and undifferentiated, *primitive* component and a somewhat differentiated, *planetary* component, with oxidation state like the highly reduced enstatite chondrites, which I suggested was identical to Mercury's complement of "lost elements". Subsequently, on the basis of that relationship, I derived expressions, as a function of the mass of planet Mercury and the mass of its core, to estimate the mass of Mercury's "lost elements", the mass of Mercury's "alloy and rock" protoplanetary core, and the mass of Mercury's gaseous protoplanet. Here, on the basis of the supposition that Mercury's complement of "lost elements" is in fact identical to the *planetary* component of ordinary-chondrite-formation, I estimate, as a function of Mercury's core mass, the total mass of ordinary chondrite matter originally present in the Solar System. Although Mercury's mass is well known, its core mass is not, being widely believed to be in the range of 70-80 percent of the planet mass. For a core mass of 75 percent, the calculated total mass of ordinary chondrite matter originally present in the Solar System amounts to $1.83 \times 10^{24}$ kg, about 5.5 times the mass of Mercury. That amount of mass is insufficient in itself to form a planet as massive as the Earth, but may have contributed significantly to the formation of Mars, as well as adding to the veneer of other planets, including the Earth. Presently, only about 0.1% of that mass remains in the asteroid belt.

## Introduction

The ordinary chondrite meteorites comprise about 80% of the meteorites that are observed falling to Earth. But their apparently great relative abundance is not necessarily a reflection of their relative importance, at least not in the manner previously thought. There is much in the scientific literature about the importance of ordinary chondrites that is not quite true, especially their perceived role as being the primary building blocks of planets.

For decades scientists have assumed that the composition of the Earth as a whole is like that of an ordinary chondrite meteorite. I have shown, however, from fundamental mass ratio relationships (Table 1), that the Earth as a whole, and particularly the endo-Earth, the inner 82% of the Earth consisting of the lower mantle and core, is not at all like an ordinary chondrite, but rather, is like an enstatite chondrite (Herndon 1980; Herndon 1982; Herndon 1998).

Ordinary chondrites have elemental abundance ratios that are similar, to within a factor of two, to many comparable element ratios in the Sun, which is an indication of their relatively simple chemical history. In fact, the often quoted equilibrium condensation model (Larimer 1967) is based upon the assumption that the minerals characteristic of ordinary chondrite meteorites condensed from a gas of solar composition and the pressure of about $10^{-5}$ bar. I have shown, however, that if the mineral assemblage characteristic of ordinary chondrites can exist in equilibrium with solar matter at all, it is at most only at a single low temperature. Moreover, the oxidized iron content of the silicates of ordinary chondrites is consistent with their formation, not only from a gas phase depleted in hydrogen by a factor of about 1000 relative to solar composition (Herndon & Suess 1977), but from a gas phase depleted somewhat in oxygen as well, as might be expected from the re-evaporation of condensed matter after separation from solar gases (Herndon 1978).

Major element fractionation among chondrites has been discussed for decades as ratios relative to Si or Mg. By expressing ratios relative to Fe, I discovered a new relationship admitting the possibility that ordinary chondrite meteorites are derived from two components, a relatively oxidized and undifferentiated, *primitive* component and a somewhat differentiated, *planetary* component, with oxidation state like the highly reduced enstatite chondrites, which I suggested is identical to Mercury's complement of "lost elements", which was presumably swept to the region of the asteroid belt by the high temperatures and/or by the violent activity during some early super-luminous solar event, such as the T-Tauri phase solar wind associated with the thermonuclear ignition of the Sun (Herndon 2004c). Subsequently, I derived expressions, as a function of the mass of planet Mercury and the mass of its core, to estimate the mass of Mercury's "lost elements" (Herndon 2004b). Here, I employ those results to estimate the total mass of ordinary chondrite matter originally present in the Solar System.

## Results and Discussion

The "standard model" of solar system formation, *i.e.*, the idea that grains condensed from diffuse nebula gases at a pressure of about $10^{-5}$ bar, and were then agglomerated into successively larger pebbles, rocks, planetesimals and, ultimately, planets, as I have shown (Herndon 2004e), is *wrong* because it would yield terrestrial planets having insufficiently massive cores, a profound contradiction to what is observed. The "standard model" of solar system formation is *wrong* because the popular underlying "equilibrium condensation" model is *wrong*, the Earth in its composition is *not* like an ordinary chondrite meteorite as frequently assumed, and condensates from nebula gases at $10^{-5}$ bar would be *too* oxidized to yield planetary cores of sufficient mass. Instead, within the framework of present knowledge, the concept of planets having formed by raining out from the central regions of hot, gaseous protoplanets, as revealed by Eucken (1944), seems to be consistent with the observational evidence (Herndon 2004d; Herndon 2004e). The nature of that condensate appears to have the same composition and state of oxidation as an enstatite chondrite meteorite.

Whole-rock major-element chondrite data, plotted as molar ratios normalized to Fe, describe three well defined lines from which the compositions of the *primitive* and *planetary* components can be obtained and from which the occurrence of the respective fractions of those two components in each ordinary chondrite can be calculated (Herndon 2004c). From the composition of the *planetary* component and from certain mass ratio relationships in enstatite chondrite matter, I derived an expression to estimate the mass of Mercury's "lost elements", contingent upon the correctness of my supposition that the complement of elements lost from Mercury's protoplanet became the planetary component of ordinary chondrite meteorites (Herndon 2004b).

The mass of the readily condensable "rock and alloy" component of the mass of the protoplanet of Mercury, $M_{Pp}$, is given by

$$M_{Pp} = M_{Mc} + M_{Mm} + M_{Lc} + M_{Lm}$$

where $M_{Mc}$ is the mass of planet Mercury's core, $M_{Mm}$ is the mass of Mercury's mantle, $M_{Lc}$ is the mass of the core-forming component of Mercury's lost elements, and $M_{Lm}$ is the mass of the mantle-forming component of Mercury's lost elements.

From (Herndon 2004b),

$$M_{Lc} = 0.3557\ M_{Mc} - 0.2492\ M_{Mn}$$

and,

$$M_{Lm} = 1.9349\ M_{Mc} - 1.3556\ M_{Mn}$$

so that the mass of Mercury's "lost elements" component, $M_{LT}$, is

$$M_{LT} = M_{Lc} + M_{Lm} = 2.2906\ M_{Mc} - 1.6048\ M_{Mn}$$

An estimate of the total mass of ordinary chondrite matter originally present in the Solar System can be obtained simply by dividing $M_{LT}$ by either the average value for the *planetary* fraction (0.255) or by the weighted average (0.238) calculated from values for individual ordinary chondrites as previously described (Herndon 2004c).

The original total mass of ordinary chondrite matter thus calculated is shown in Fig. 1 as a function of the relative mass of Mercury's core. Although the mass of planet Mercury, $3.3022 \times 10^{23}$ kg, is well known, considerable uncertainty presently exists as to the precise mass of its core. There is widespread belief that Mercury's core accounts for 70-80 percent of the planet mass. The great uncertainty in core mass arises primarily from the uncertainty in Mercury's moment of inertia, which will be measured during NASA's Messenger mission.

For a core comprising 75% of Mercury's mass, the calculated total original mass of ordinary chondrite matter in the Solar System amounts to $1.83 \times 10^{24}$ kg, about 5.5 times the mass of Mercury. That amount of mass is insufficient in itself to form a planet as massive as the Earth, but may have contributed significantly to the formation of Mars, as well as adding to the veneer of other planets, including the Earth. Presently, only about 0.1% of that mass remains in the asteroid belt (Cox 2000).

The calculations made in this paper are based upon the supposition that Mercury's "lost elements" component is one of the two components from which the ordinary chondrites were formed. Although circumstantial evidence supports that supposition (Herndon 2004b), the challenge ahead is to make discoveries or to discover fundamental quantitative relationships which will change that supposition into fact.

# References

Cox, A. N. 2000 *Allen's Astrophysical Quantities*. New York: AIP and Springer-Verlag.

Eucken, A. 1944 Physikalisch-chemische Betrachtungen ueber die frueheste Entwicklungsgeschichte der Erde. *Nachr. Akad. Wiss. Goettingen, Math.-Kl.*, 1-25.

Herndon, J. M. 1978 Reevaporation of condensed matter during the formation of the solar system. *Proc. R. Soc. Lond* **A363**, 283-288.

Herndon, J. M. 1980 The chemical composition of the interior shells of the Earth. *Proc. R. Soc. Lond* **A372**, 149-154.

Herndon, J. M. 1982 The object at the centre of the Earth. *Naturwissenschaften* **69**, 34-37.

Herndon, J. M. 1998 Composition of the deep interior of the earth: divergent geophysical development with fundamentally different geophysical implications. *Phys. Earth Plan. Inter* **105**, 1-4.

Herndon, J. M. 2004a Background for terrestrial antineutrino investigations: scientific basis of knowledge on the composition of the deep interior of the Earth. *arXiv:hep-ph/0407149 13 July 2004*.

Herndon, J. M. 2004b Mercury's protoplanetary mass. *arXiv:astro-ph/0410009 1 Oct 2004*.

Herndon, J. M. 2004c Ordinary chondrite formation from two components: implied connection to planet Mercury. *arXiv:astro-ph/0405298 15 May 2004*.

Herndon, J. M. 2004d Protoplanetary Earth formation: further evidence and geophysical implications. *arXiv:astro-ph/0408539 30 Aug 2004*.

Herndon, J. M. 2004e Solar System formation deduced from observations of matter. *arXiv:astro-ph/0408151 9 Aug 2004*.

Herndon, J. M. & Suess, H. E. 1977 Can the ordinary chondrites have condensed from a gas phase? *Geochim. Cosmochim. Acta* **41**, 233-236.

Jarosewich, E. 1990 Chemical analyses of meteorites: A compilation of stony and iron meteorite analyses. *Meteoritics* **25**, 323-337.

Larimer, J. W. 1967 Chemical fractionation in meteorites I, Condensation of the elements. *Geochim. Cosmochim. Acta* **31**, 1215-1238.

Norton, O. R. 2002 *The Cambridge Encyclopedia of Meteorites*. Cambridge: Cambridge University Press.

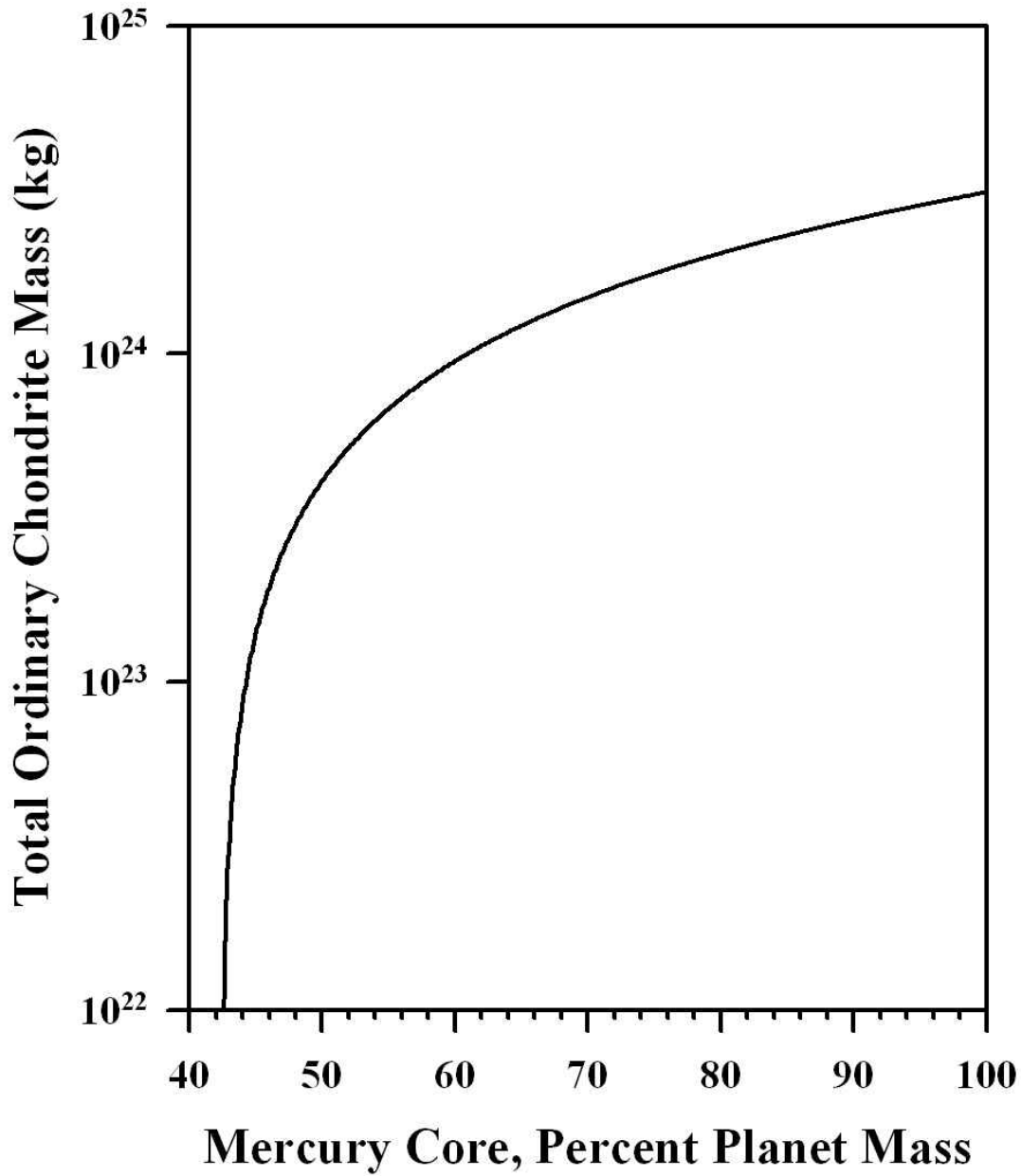

**Figure 1**. Lacking knowledge of the precise value for the mass of Mercury's core, the calculated total mass of ordinary chondrite matter originally present in the Solar System is shown for a range of masses. The curve was calculated by dividing the mass of Mercury's "lost element" component by the weighted average *planetary* fraction from data on 157 individual ordinary chondrites (Jarosewich 1990) and weighted by the relative proportion of H, L, and LL ordinary chondrites observed to fall (Norton 2002).

**Table 1**. Fundamental mass ratio comparison between the endo-Earth (core plus lower mantle) and the Abee enstatite chondrite (Herndon 2004a).

| Fundamental Earth Ratio | Earth Ratio Value | Abee Ratio Value |
|---|---|---|
| lower mantle mass to total core mass | 1.49 | 1.43 |
| inner core mass to total core mass | 0.052 | *theoretical* 0.052 if $Ni_3Si$ 0.057 if $Ni_2Si$ |
| inner core mass to (lower mantle+core) mass | 0.021 | 0.021 |
| core mean atomic mass | 48 | 48 |
| core mean atomic number | 23 | 23 |